\documentclass[12pt]{iopart}
\usepackage{graphicx}
\usepackage[normalem]{ulem}
\usepackage{bm}
\newcommand{\ket}[1]{\vert#1\rangle}

\usepackage{color}

\begin{document}
\title{An integrated processor for photonic quantum states using a broadband light-matter interface}
\author{E.~Saglamyurek, N.~Sinclair, J.~A.~Slater, K.~Heshami, D.~Oblak and W.~Tittel}
\address{Institute for Quantum Science and Technology, Department of Physics \& Astronomy, University of Calgary, 2500 University Drive
  NW, Calgary, Alberta T2N 1N4, Canada.}
\ead{esaglamy@ucalgary.ca}

\begin{abstract}
Faithful storage and coherent manipulation of quantum optical pulses are key for long distance quantum communications and quantum computing. Combining these functions in a light-matter interface that can be integrated on-chip with other photonic quantum technologies, e.g. sources of entangled photons, is an important step towards these applications. To date there have only been a few demonstrations of coherent pulse manipulation utilizing optical storage devices compatible with quantum states, and that only in atomic gas media (making integration difficult) and with limited capabilities. Here we describe how a broadband waveguide quantum memory based on the Atomic Frequency Comb (AFC) protocol can be used as a programmable processor for essentially arbitrary spectral and temporal manipulations of individual quantum optical pulses.~Using weak coherent optical pulses at the few photon level, we experimentally demonstrate sequencing, time-to-frequency multiplexing and demultiplexing, splitting, interfering, temporal and spectral filtering, compressing and stretching as well as selective delaying. Our integrated light-matter interface offers high-rate, robust and easily configurable manipulation of quantum optical pulses and brings fully practical optical quantum devices one step closer to reality. Furthermore, as the AFC protocol is suitable for storage of intense light pulses, our processor may also find applications in classical communications.
\end{abstract}
\maketitle
\section{Introduction}
\begin{figure}[t]
\begin{center}
\includegraphics[width=1\columnwidth,angle=0]{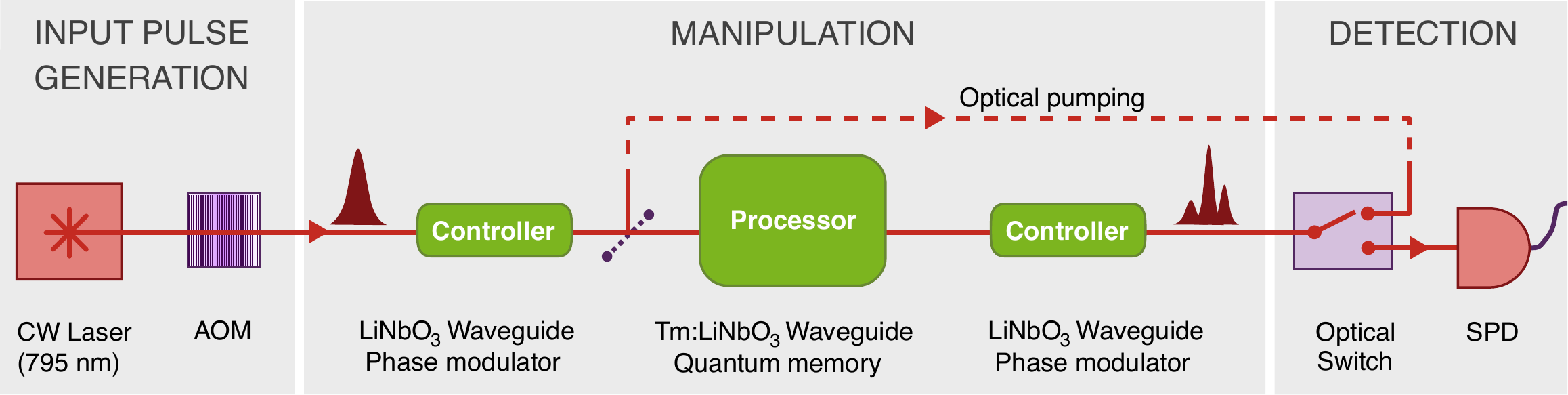}
\caption[Simplified diagram of experimental setup for optical pulse manipulation based on AFC processor.]{\bf Simplified diagram of experimental setup for optical pulse manipulation based on AFC processor.}
\label{pulsemanipsetup}
\end{center}
\end{figure}

Since the invention of the laser, a lot of effort has been put into manipulating optical pulses for applications in science and engineering. One important application is photon-based quantum information processing, which promises unbreakable secret key distribution and unprecedented computational power \cite{Gisin_02,Obrien_09}. The realization of these applications requires coherently storing as well as manipulating quantum optical pulses in order to process and distribute quantum information \cite{Obrien_09, Lvovsky_09, Sangouard_11}. Combining storage and manipulation in a light-matter interface that can be integrated on-chip with other photonic components reduces the complexity and thus facilitates the development of future quantum technologies \cite{Obrien_09, Sohler_08, Tanzilli_12, Obrien_12}. To date there have only been a few investigations that employ a quantum storage device for coherent optical pulse manipulation \cite{Hosseini_09, Benson_10, Novikova_11, Sparkes_12, Walmsley_12}. However, these demonstrations feature various intrinsic limitations that impact future use in practical settings. First, all previous demonstrations rely on atomic vapor, making integration difficult. Second, most show limited processing capabilities, which restricts their use to a small number of specific applications \cite{Benson_10, Novikova_11, Walmsley_12}. Third, the bandwidths in \cite{Hosseini_09, Benson_10, Novikova_11, Sparkes_12} were at most a few MHz, which, accordingly, constraints the minimum duration of the processed pulses to a few microsecond. Finally, the number of simultaneously storable (and hence processable) qubits in the quantum storage devices considered before is severely limited \cite{Hosseini_09, Benson_10, Novikova_11, Sparkes_12, Walmsley_12, Nunn_08}.

In this paper, we propose and demonstrate a universal, large-bandwidth and multimode approach to temporal and spectral manipulation of individual quantum optical pulses -- it relies on introducing a broadband quantum memory based on a Tm:LiNbO$_3$ waveguide and the Atomic Frequency Comb (AFC) protocol \cite{Saglamyurek_11, Saglamyurek_12} between two LiNbO$_3$ waveguide phase modulators that serve as variable frequency shifters using the serrodyne frequency-translation technique \cite{Serrodyne_1} (see Appendix A7 for more details). Reversibly mapping nano-second long pulses of light onto different spectral sections of a multi-section AFC (where each spectral section is programmed to perform a different task), we demonstrate sequencing, time-to-frequency multiplexing and demultiplexing, splitting, interfering, temporal and spectral filtering, compressing and stretching as well as on-demand selective delaying. While the input pulses currently contain on average around 5--40 photons, our approach straightforwardly applies to the manipulation of individual photons or members of entangled photon pairs, as well as to pulses of light used in classical communications. 

\section{Experiment} \label{experiment}
\begin{figure}
\begin{center}
\includegraphics[width=1\columnwidth,angle=0]{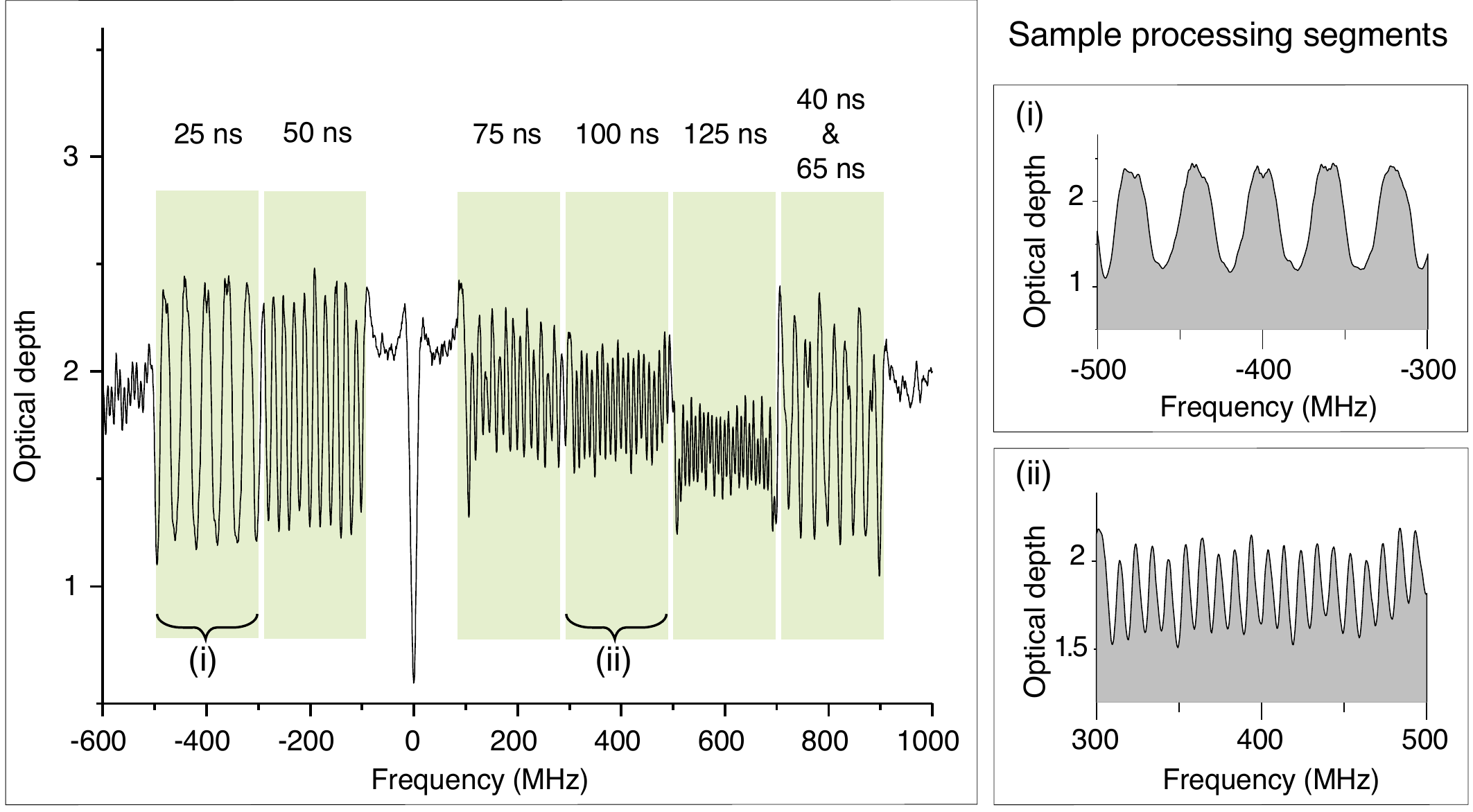}
\end{center}
\caption[Example of a programmed AFC memory as a processor used in the experiments.]{\textbf{Example of a programmed AFC memory used as a processor in the experiments}. See main text in Section~\ref{experiment} for details.}
\label{processor}
\end{figure}
Our experimental setup, sketched in Fig.~\ref{pulsemanipsetup}, consists of three main blocks: pulse generation, pulse manipulation, and detection. The pulse generation block features a continuous wave (CW) laser centered at 795.5 nm wavelength and an acousto-optic modulator (AOM) that is used to carve 12 ns duration pulses (measured at Full Width Half Maximum, FWHM) from the CW laser light. These pulses are heavily attenuated down to a mean photon number of $5-40$ photons. The pulse manipulation block is composed of a Tm:LiNbO$_3$ waveguide memory (processor) maintained at 3.5 K, and two commercial LiNbO$_3$ waveguide phase modulators (controllers) located at the input and output of the memory and kept at room temperature. We note that it may be possible to combine these three elements on a single, integrated, cryogenically cooled photonic circuit, allowing a more compact setup. The detection block contains a silicon avalanche photo diode (Si-APD) single photon detector (SPD), an optical switch that is used to protect the detector from optical pump light while preparing the AFC, and a monolithic Fabry-Perot (FP) filter (not shown in the figure) used to demonstrate the ability to control the spectra of the retrieved pulses from the memory unit (this is further explained in Appendix A5). Each experimental cycle comprises three stages: AFC programming through frequency selective optical pumping (persistent spectral hole burning) for which the frequency of the CW laser light is controlled using the phase modulator placed before the memory, waiting to avoid noise from decaying atoms that were excited during the programming of the AFC, and sending, manipulating and detecting few-photon pulses. The durations of these stages are 3 ms, 2.2 ms and 5 ms, respectively. The efficiency of our AFC processor is currently limited to around 1$\%$. As is further discussed in Section~\ref{discussion}, this is  mainly due to inefficient optical pumping during the programming stage of the AFC. For more details about the implementation, material, AFC preparation and measurements see Appendix A1-A5. 

In a standard AFC memory \cite{Riedmatten_08, Afzelius_09}, an inhomogeneously broadened absorption profile is tailored into a series of equally spaced absorption peaks. Photons mapped onto this spectral feature are re-emitted as so-called echoes after a time $t_{storage}=1/{\Delta}$, and with a phase shift of $\phi=2 \pi \frac{\Delta_{0}}{\Delta}$, where $\Delta$ is the spacing between the peaks, and $\Delta_{0}$ is the detuning between the carrier frequency of the input photons and the frequency of one of the peaks \footnote{We note that, in the case of backward recall and assuming zero detuning, the echo picks up an additional phase of $\pi$ with respect to the input. However, this is not the case for the forwards propagating echo as implemented in our experiments.}. The retrieval process can approach unit efficiency provided that certain phase matching conditions are satisfied and variables such as AFC finesse (i.e. the ratio of the peak spacing to the peak width) and optical depth are optimum. The implementation of the AFC protocol in cryogenically cooled rare-earth ion doped crystals has already shown great promise as a quantum memory for quantum information processing applications. Examples include high retrieval fidelity \cite{Guo_12}, high efficiency \cite{Sabooni_13}, large bandwidth \cite{Saglamyurek_11,Saglamyurek_12}, large temporal and spectral multimode storage capacity \cite{Usmani_10, Thierry_11, Sinclair_13}, the possibility to store time-bin and polarization qubits~\cite{Saglamyurek_12, Guo_12, Gundogan_12, Clausen_12} plus members of entangled photon pairs \cite{Saglamyurek_11, Clausen_11} and, recently, the teleportation of photonic quantum states into a crystal \cite{Felix_14}.    

When the AFC is used for processing tasks, as proposed in this study, the peak spacing in different frequency intervals is typically set to different values, resulting in storage (re-emission) times that vary as a function of frequency. An example of an AFC memory programmed as a processor is shown in Fig.~\ref{processor}. In this example, the entire bandwidth is divided into six AFC segments, each having 200 MHz bandwidth. In each AFC segment, the peak spacings are set differently, yielding storage times between 25 ns and 125 ns with 25 ns increments. The right-most segment, prepared by superimposing two AFCs, is used to divide a single input pulse into two temporal modes, stored for 40 ns and 65 ns, respectively. Using this single re-configurable AFC processor, multiple processing tasks can be performed, as shown in the following~\footnote{Note that the retrieval efficiency of each AFC segment, if programmed for different storage times, is typically different. However, as explained in Appendix~A3, the efficiencies can be made equal at the expense of lowering the overall efficiency. This was done for the measurements depicted in Figures~3--6.}.

\begin{figure}[t!]
\begin{center}
\includegraphics[width=1\columnwidth,angle=0]{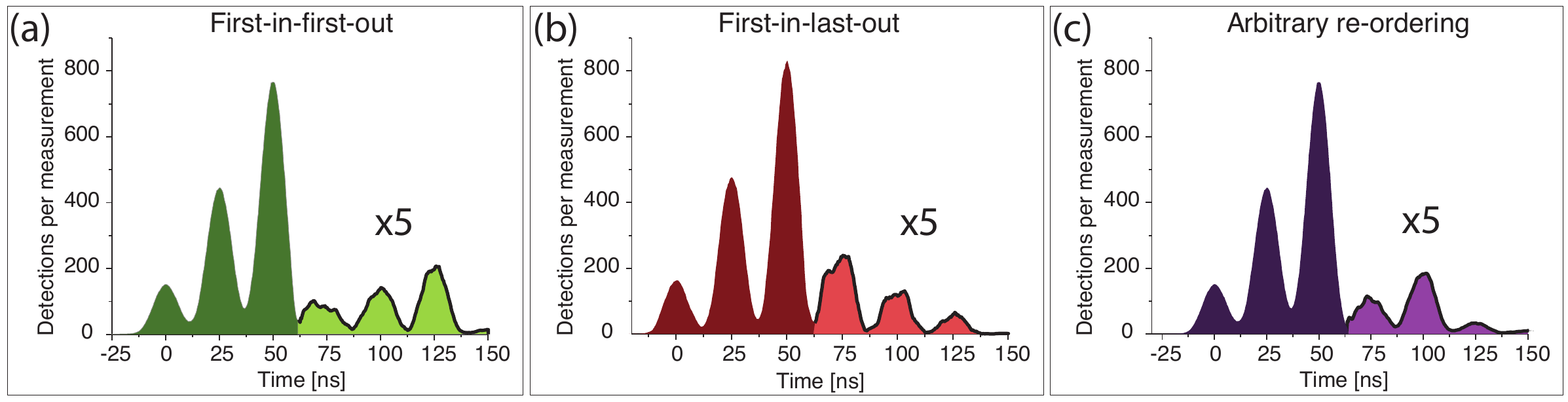}
\end{center}
\caption[Demonstration of pulse sequencing]{\textbf{Demonstration of pulse sequencing:} An input pulse sequence consisting of three pulses separated by 25 ns is generated using the AOM. The pulses are prepared with 12 ns duration and the same carrier frequency but with different amplitudes. \textbf{(a)} The input pulses are all mapped onto the same AFC segment at +200 MHz detuning using the input phase modulator. After 75 ns storage they are retrieved in the same order (First-in-first-out). \textbf{(b)} The first, second and last pulses are mapped onto the AFC segments that yield 125 ns, 75 ns and 25 ns storage time, respectively. This results in the pulses being recalled in a time-reversed order with respect to the input sequence (First-in-last-out). In the measurements, the retrieval efficiency for each storage time is equalized (at the expense of lowering the overall efficiency) by choosing an appropriate magnetic field and pump intensity during the programming stage. \textbf{(c)} Similarly, by applying appropriate frequency shifts to each pulse, arbitrary time re-ordering can be obtained.}
\label{pulsereordering}
\end{figure}
\subsection{\textbf{Pulse Sequencing}} \label{pulsereorder}
\begin{figure}[t!]
\begin{center}
\includegraphics[width=0.80\columnwidth,angle=0]{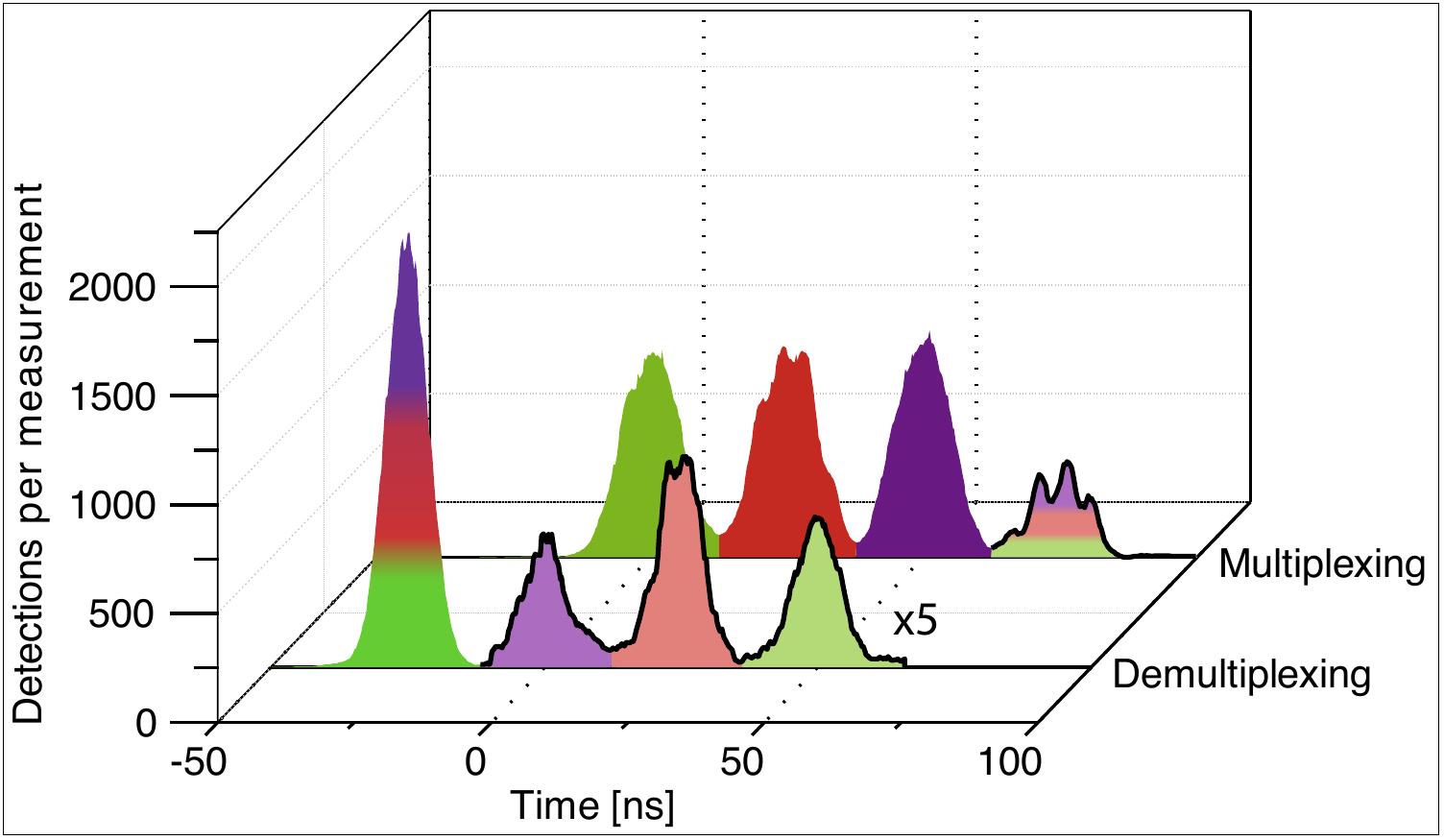}
\end{center}
\caption[Demonstration of time-to-frequency multiplexing and demultiplexing.]{\textbf{Demonstration of time-to-frequency multiplexing and demultiplexing:} 
A pulse composed of three distinct frequency modes at -400 MHz, -200 MHz and +200 MHz (represented by the three horizontal colour bands in the pulse) detunings is generated using the input phase modulator driven by a sinusoidal waveform. This multimode pulse is sent to the programmed AFC (Fig.~\ref{processor}) and each frequency mode is mapped onto the corresponding AFC segment. This results in each frequency mode being retrieved at a different time, as determined by the programming of the respective AFC segment. This is illustrated in the front trace (Demultiplexing) with different colours representing the different frequency components. Next, three 25 ns-separated input pulses are prepared with center frequencies that map onto the AFC segments with +200 MHz, -200 MHz and -400 MHz detuning, respectively. This mapping allows the three recalled pulses to be merged in the same temporal mode, as shown in the back trace (Multiplexing).}
\label{multiplexing}
\end{figure}
The first example is the re-ordering of pulses in a pulse sequence, which plays an important role in synchronizing and randomly accessing quantum information in quantum repeaters and linear optics quantum computers \cite{Hosseini_09}. In the standard AFC storage scheme, one is restricted to storing pulses without changing their order, as shown in Fig.~\ref{pulsereordering}a \footnote{Here and in the reminder of this paper, the recalled pulses (echoes) are identified by a thick outline while the transmitted parts of the incoming pulses, arising from the non-unit absorption probability of the incoming pulses, are shown in a darker color, with no outline and, for the most part, will be located at the origin of the time axis.}. To re-order the pulses in this sequence, the controller at the AFC memory input applies appropriate frequency shift to each pulse such that the pulses are mapped onto different AFC segments, resulting in the retrieved pulse sequence being reversed or arbitrarily re-ordered, as demonstrated in Fig.~\ref{pulsereordering}b and c, respectively. Moreover, the center frequency of each pulse retrieved from the memory unit can be set back to its original value by the output controller, resulting in a pure time-domain manipulation (see Appendix A5).

\subsection{\textbf{Time-to-Frequency Multiplexing and Demultiplexing}}\label{FTMD}

The next demonstration is time-to-frequency multiplexing and demultiplexing of pulses, which is of potential interest for multiplexed quantum networks. When input pulses occupying several frequency modes and the same temporal mode are mapped onto the programmed AFC processor in Fig.~\ref{processor}, they are retrieved at different times (demultiplexing) as shown in the front trace of Fig.~\ref{multiplexing}. In the same way, input pulses occupying different temporal modes can be merged into the same temporal mode (multiplexing) by applying the appropriate frequency shift to each input mode as demonstrated in the back trace of Fig.~\ref{multiplexing}. 
\begin{figure}[t!]
\begin{center}
\includegraphics[width=0.50\columnwidth,angle=0]{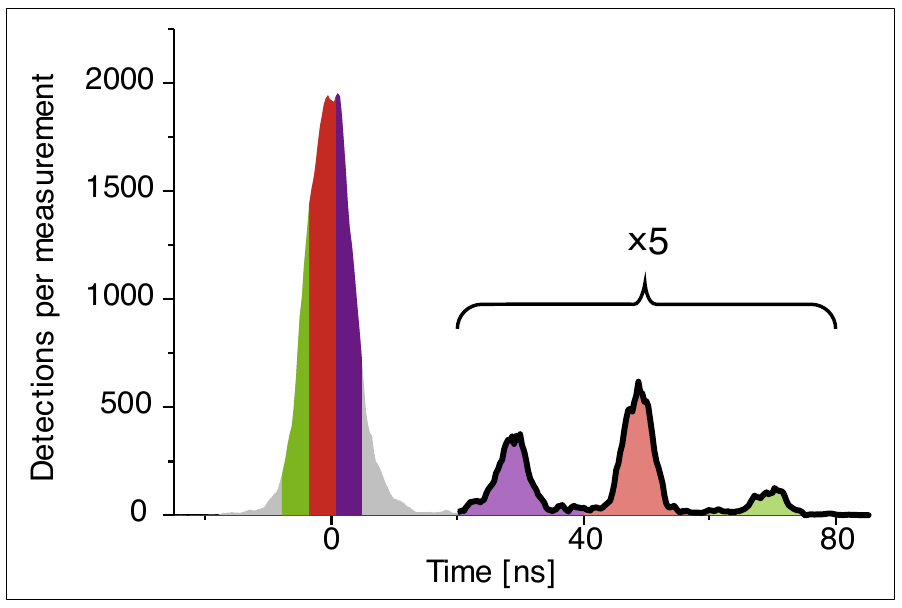} 
\end{center}
\caption[Pulse splitting]{\textbf{Pulse splitting:} Three separate temporal portions of an input pulse, indicated by the green, red and purple coloured bands are mapped to three different segments of the programmed AFC, leading to 75 ns, 50 ns and 25 ns storage times, respectively. After processing in the AFC, three spectro-temporal modes emerge as shown in the figure.}
\label{pulsesplit}
\end{figure}
\subsection{\textbf{Pulse Splitting}}

Splitting a pulse into separate pulses in various spectral and temporal modes is another manipulation that is possible using our processor; it allows generating high-dimensional quantum states. To demonstrate this feature, three temporal portions of an input pulse are mapped onto different AFC processor segments using the input controller. Consequently, each portion of the original pulse is retrieved in a different spectro-temporal mode, as shown in Fig.~\ref{pulsesplit}. We note that it is also possible to recover the input pulse by re-processing each generated component in a subsequent system.

\subsection{\textbf{Manipulating Time-Bin and Frequency-Bin Qubit States}}

\begin{figure}[ht!]
\begin{center}
\includegraphics[width=1\columnwidth,angle=0]{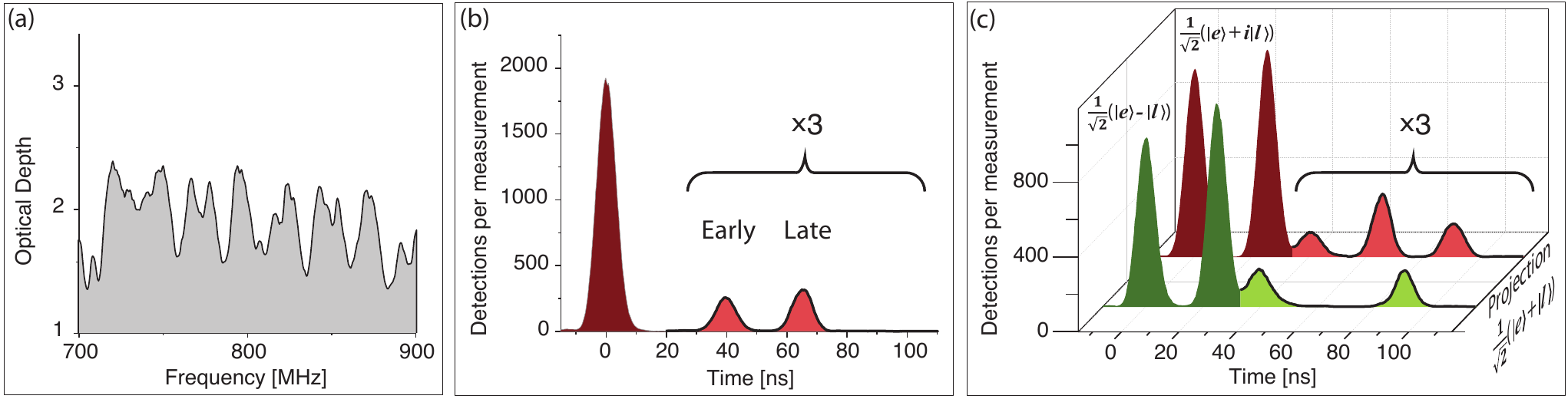} 
\end{center}
\caption[Generating and manipulating time-bin qubit states using programmed AFC processor]{\textbf{Generating and Manipulating Time-Bin Quantum States:} \textbf{(a)} The last segment of the programmed AFC at +800 MHz detuning (Fig.~\ref{processor}) is prepared by superimposing two AFCs (double AFC) with two different comb spacings, leading to 40 ns and 65 ns storage times. \textbf{(b)} An input photon, occupying a well-defined temporal mode, is mapped to the double AFC. It is re-emitted in a superposition of two temporal modes (time bins), referred to as ``early'' and ``late''.~\textbf{(c)} Storing a superposition time-bin qubit state with temporal modes separated by 25 ns (equal to the difference between the re-emission times of the superimposed AFCs) leads to the early and late input modes being overlapped in a central temporal mode after recall (see text for details). This allows  performing any projection measurement onto time-bin qubit states. In the measurements represented in (c), input states $\ket{\psi}=\frac{1}{\sqrt{2}}\left(\ket{e}-\ket{l}\right)$ (front trace) and $\ket{\psi}=\frac{1}{\sqrt{2}}\left(\ket{e}+i\ket{l}\right)$ (back trace), where $\ket{e}$ and $\ket{l}$ denote the quantum state of a photon occupying the early and late temporal mode, respectively, are stored in a double AFC that projects onto $\ket{\psi}=\frac{1}{\sqrt{2}}\left(\ket{e}+\ket{l}\right)$. This results in destructive and (partially) constructive interference in the central temporal mode, respectively.}
\label{doublegrating}
\end{figure}

It is straightforward to generate and manipulate time-bin qubits using our integrated processing unit, as previously demonstrated in~\cite{Saglamyurek_12, Riedmatten_08, Usmani_10, Clausen_11}. In the programmed AFC processor shown in Fig.~\ref{processor}, the last segment at +800 MHz detuning is prepared by superimposing two AFCs (double AFC) with different comb tooth spacings (see Fig.~\ref{doublegrating}a). When an input pulse in a well defined temporal mode is mapped onto this AFC segment, it is re-emitted in a superposition of two temporal modes with a relative phase of $\delta \phi=\phi_1-\phi_2$, where $\phi_1$ and $\phi_2$ are the phase shifts introduced by each comb of the double AFC. Thus, this AFC processor segment can be used to generate any time-bin qubit state, as exemplified in Fig.~\ref {doublegrating}b. 

If a single photon pulse in a superposition of two temporal modes is mapped onto the AFC segment, the photon will be re-emitted, in general, in a superposition of four temporal modes, each defined by the re-emission times of the combs and the separation of the input temporal modes. In this case, adjusting the difference between the emission times to the temporal separation of the two input modes leads to overlap between the two central temporal modes into a single mode, as shown in Fig.~\ref{doublegrating}c. In this way, it is possible to project a time-bin qubit onto an arbitrary state \cite{Riedmatten_08}, which resembles the use of an imbalanced Mach-Zender interferometer. Furthermore, applying the same idea used for time-to-frequency multiplexing (demonstrated in Sec. \ref{FTMD}), our system can easily convert a time-bin qubit to a frequency-bin qubit, which can be further manipulated using the output controller. These schemes could be useful for quantum communication protocols relying on frequency and time-bin encoding.  

\subsection{\textbf{Arbitrary Temporal and Spectral Filtering}} \label{filter}
\begin{figure}[t!]
\begin{center}
\includegraphics[width=1\columnwidth,angle=0]{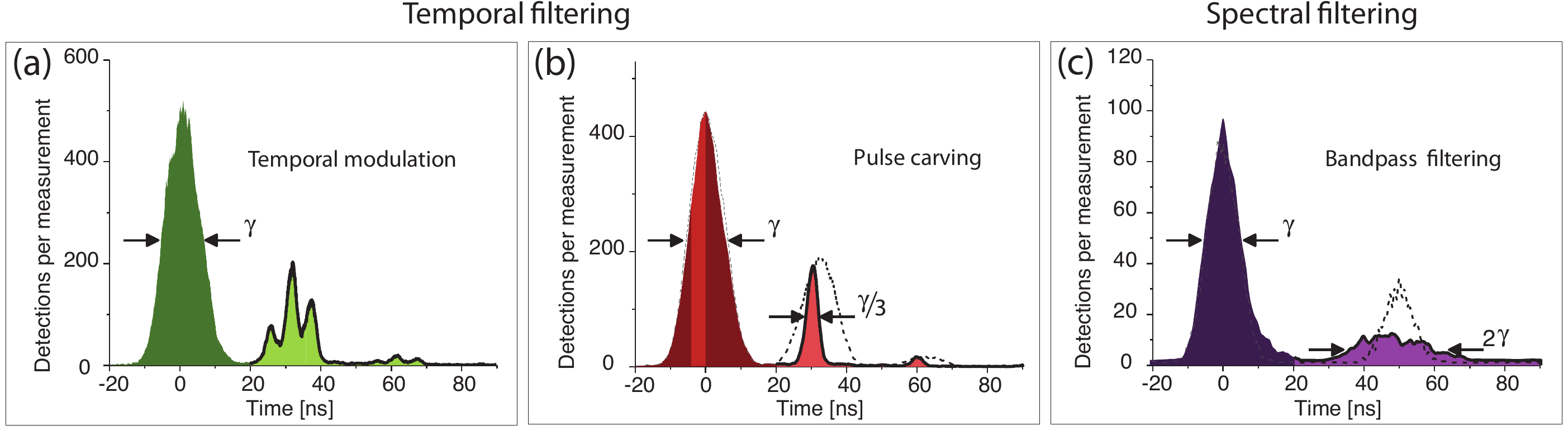} 
\end{center}
\caption[Arbitrary temporal and spectral filtering]{\textbf{Arbitrary temporal and spectral filtering:} \textbf{(a)} The input phase modulator maps three selected 3 ns-long temporal portions of a 12 ns long input pulse onto an AFC segment prepared with 600 MHz bandwidth, 33 MHz peak spacing, and centered at +700 MHz detuning. This allows generating a temporally modulated pulse as seen in the figure. \textbf{(b)} The input phase modulator maps a narrow temporal portion of an input pulse onto the center of the AFC segment. This results in an echo almost three times shorter than the input pulse. The echo pulse, shown with the dashed line is obtained if the entire pulse is mapped to the same AFC. $\gamma$ represents the duration of the input and transmitted pulse at FWHM. \textbf{(c)} An input pulse with nearly 70 MHz bandwidth is mapped onto an AFC segment prepared with 600 MHz bandwidth and retrieved after 50 ns (dashed line). An input pulse with the same bandwidth is mapped to an AFC with a bandwidth decreased to about 40 MHz. In this case the AFC acts as bandpass filter, yielding an output pulse with nearly twice the duration.} 
\label{pulseshaping}
\end{figure}

Our processor can also be used as a reconfigurable temporal and spectral filter. For example, this would allow tailoring single photons or matching the spectral and temporal properties of photons produced from independent sources for applications relying on two-photon interference. As opposed to the task described in the next section, filtering obviously leads to loss of photons. To demonstrate the temporal shaping capability, an AFC having a bandwidth of 600 MHz and centered at +700 MHz detuning is programmed. A 12 ns long input pulse is sent to the processing unit. If the input controller is not engaged, it remains at 0 MHz detuning and the pulse is directly transmitted as it has no spectral overlap with the AFC. However, if the input controller is engaged, the selected temporal portion (or portions) of the incoming pulse is (are) mapped onto the AFC, resulting in a temporally shaped recalled pulse as shown in Fig.~\ref{pulseshaping}a. Using this idea, it is also possible to carve a narrow temporal portion from an input pulse, i.e. to generate a short pulse from a long one, as demonstrated in Fig.~\ref{pulseshaping}b. Furthermore, the capability of continuously adjusting the serrodyne modulation efficiency (see Appendix A7) allows for arbitrary modulation of the temporal shape and intensity of an input pulse, which turns our implementation into an arbitrary optical waveform generator \cite{Babbitt_02}.   

To demonstrate the spectral filtering capability, the bandwidth of the AFC is decreased to about 40 MHz. An input pulse with 70 MHz bandwidth is mapped onto this AFC using the input controller. The portion of the pulse spectrum that lies outside the AFC is filtered out (by being partially directly transmitted and absorbed) and the recalled pulse is emitted with smaller bandwidth and thus longer duration, see Fig.~\ref{pulseshaping}c. 

\begin{figure}[t!]
\begin{center}
\includegraphics[width=0.9\columnwidth,angle=0]{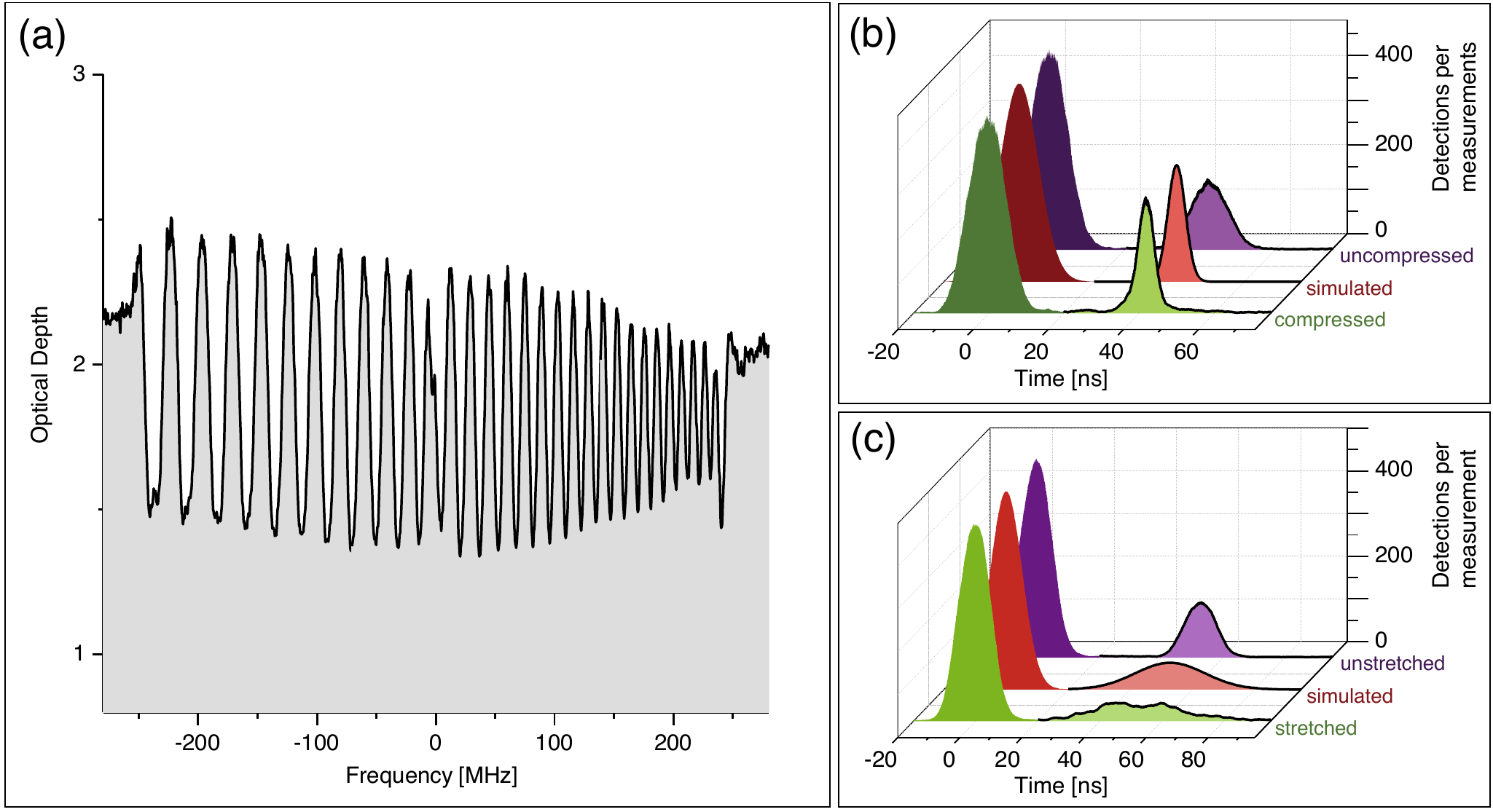} 
\end{center}
\caption[Compressing and stretching of pulses using an AFC with variable peak spacing.]{\textbf{Demonstration of compressing and stretching of pulses using an AFC with linearly- decreasing peak spacing:} \textbf{(a)} A 500 MHz-wide AFC with linearly decreasing peak spacing from 30 MHz to 10 MHz, yielding storage times between 33 ns (at the lower frequency end of the AFC) and 100 ns (at the upper frequency end of the AFC, respectively), and thus an approximate value of $\mu=0.14$ ns/MHz. \textbf{(b)} Using the input phase modulator, the frequency of an incoming pulse is linearly chirped from high to low at a rate of 8.3~MHz/ns, yielding $\mu r =1.16$ (close to the optimum condition $\mu r=1$). When this pulse is mapped onto the programmed AFC, a compressed pulse is retrieved with a compression factor of nearly 2.4, as shown in the front trace. The middle trace depicts the simulated result after scaling its intensity with respect to the experimentally obtained echo (see Appendix 6)-- the temporal shapes are in good agreement. The back trace shows the retrieved pulse after storage using a standard AFC with equal tooth spacing. Note that the areas under the experimentally obtained curves for the standard and compressed echoes are nearly equal. \textbf{(c)} Pulse stretching is performed by applying a chirp with the same rate but different sign (i.e. from low to high frequency) and different start frequency. The resulting stretched pulse, the corresponding simulation and the re-emitted pulse from a standard AFC are shown in the front, middle and back traces, respectively.}
\label{compression1}
\end{figure}

\subsection {\textbf{Pulse Compressing and Stretching}} \label{compress}
\begin{figure}[t!]
\begin{center}
\includegraphics[width=0.55\columnwidth,angle=0]{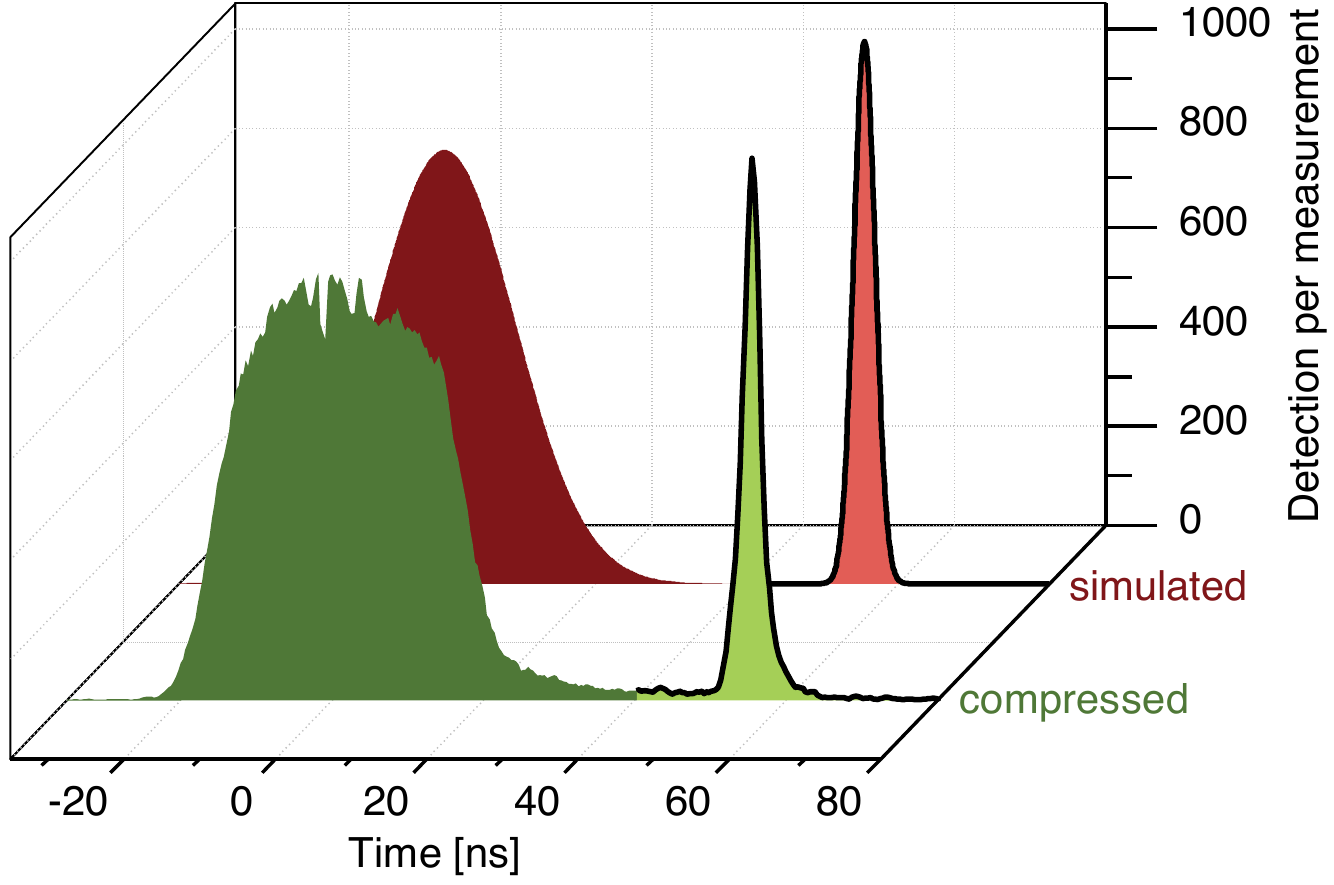}
\end{center}
\caption[Achieving high compression factors]{\textbf{Achieving high compression factors:} A 2 GHz-wide AFC with linearly decreasing peak spacing from 30 MHz to 10 MHz (equivalent to a storage time gradient $\mu=0.035$) is programmed. The frequency of an input pulse with approximately 30 ns duration is chirped from high to low at a rate of 32.9 MHz/ns, yielding $\mu r=1.15$, and mapped onto the programmed AFC. This results in a 2.6 ns long (FWHM) echo pulse and a compression factor close to 10, as shown in the front trace. The back trace shows the simulated result assuming an input pulse having a Gaussian envelope and a similar duration to that of the actual input.} \label{highcompression} 
\end{figure}

Compressing and stretching quantum optical pulses is potentially useful for increasing quantum data rates and matching bandwidths of photons to those of quantum  memories \cite{Hosseini_09, Sparkes_12, Moiseev_10}. An AFC memory is well suited for this task since it can be tailored to feature almost arbitrary chromatic dispersion, i.e. tailored in such a way that different frequency components of an input pulse are retrieved at different times. To this end, we program an AFC whose peak separation is linearly decreasing across its entire bandwidth (Fig.~\ref{compression1}a). To compress an input pulse, the input controller chirps the frequency of the  pulse linearly from high to low, which disperses the spectrum of the pulse.  When this pulse is mapped onto the AFC processor, the pulse front is stored longer than the end of the pulse, leading, under certain conditions, to the retrieval of a temporally compressed pulse as demonstrated in Figs.~\ref{compression1}b and ~\ref{highcompression}. Note that the output controller can be used to remove the frequency chirp in the retrieved pulse by applying a reversed linear chirp with appropriately chosen slope. 
 
For an input pulse having a duration of $\tau_{in}$ and an AFC prepared with linearly varying peak spacing, the compression factor $\kappa$ (defined to be the ratio of the duration of the input pulse to the duration of the compressed output pulse) is determined by two parameters: the spectral width of the chirped pulse, $B=\tau_{in} r$, which is controlled by the applied chirp rate, $r$, as well as the gradient of the programmed storage times  $\mu=\frac {\delta T} {B}$. Here, $\delta T$ is the difference between the storage times introduced by the AFC for the upper frequency end and the lower frequency end of the chirped pulse (see Appendix~A6 for the analysis). Note that we assume that the AFC width is at least as large as $B$ so that the entire chirped pulse is absorbed. For maximum compression of the input pulse, $\tau_{in}$ and $\delta T$ have to be equal, which is equivalent to satisfying the condition $\mu r=1$ (this can be achieved using an appropriate chirp-rate setting). In this case, the duration of the output pulse is $1\slash B$ \cite{Babbitt_02}, and the compression factor is given by 

\begin{equation}
\label{compressionfactor}
\kappa=\frac{\tau_{in}}{1\slash B}=\delta T\cdot B=\mu B^{2}.
\end{equation}
\noindent This equation follows straightforwardly from the definition for the storage time gradient $\mu$. Hence, we find that, for a given storage material, the compression factor is limited by the maximally achievable storage time (which determines $\delta T$) and the storage bandwidth $B$ (i.e. the time-bandwidth product). For instance, in our case, we have $\delta T_{max}\approx$ 100 ns and $B_{max}\approx$ 10 GHz, which allows in principle a compression factor of  $\kappa_{max}\approx$1000. We note that compression not only occurs if $\mu r=1$, but, more generally, for $0<\mu r<2$ (for more information see Appendix 6).

Pulse stretching can be performed in a similar way by applying the input frequency chirp from low to high. When the chirped pulse is mapped onto an AFC with decreasing peak spacing, the end of the pulse is stored longer than the front, leading to a temporally stretched pulse of duration $\tau_{out}\approx\tau_{in}+\delta T=\tau_{in}+\mu B$ (an example is shown in Fig.~\ref{compression1}c). Hence, achieving large stretching requires programming an AFC with large $\mu$, which, as before, is limited by the longest achievable storage time (i.e. the smallest achievable peak spacing of the AFC). In addition, the output pulse width increases with the bandwidth $B$ of the chirped input pulse. 

To finish this section, let us emphasize that the described pulse compression and decompression is different from the filtering process discussed in section 2.5 as, in principle, this approach can be carried out without introducing loss.

\subsection{\textbf{Selective Delaying}} \label{ondemand}
\begin{figure}[t!]
\begin{center}
\includegraphics[width=0.9\columnwidth,angle=0]{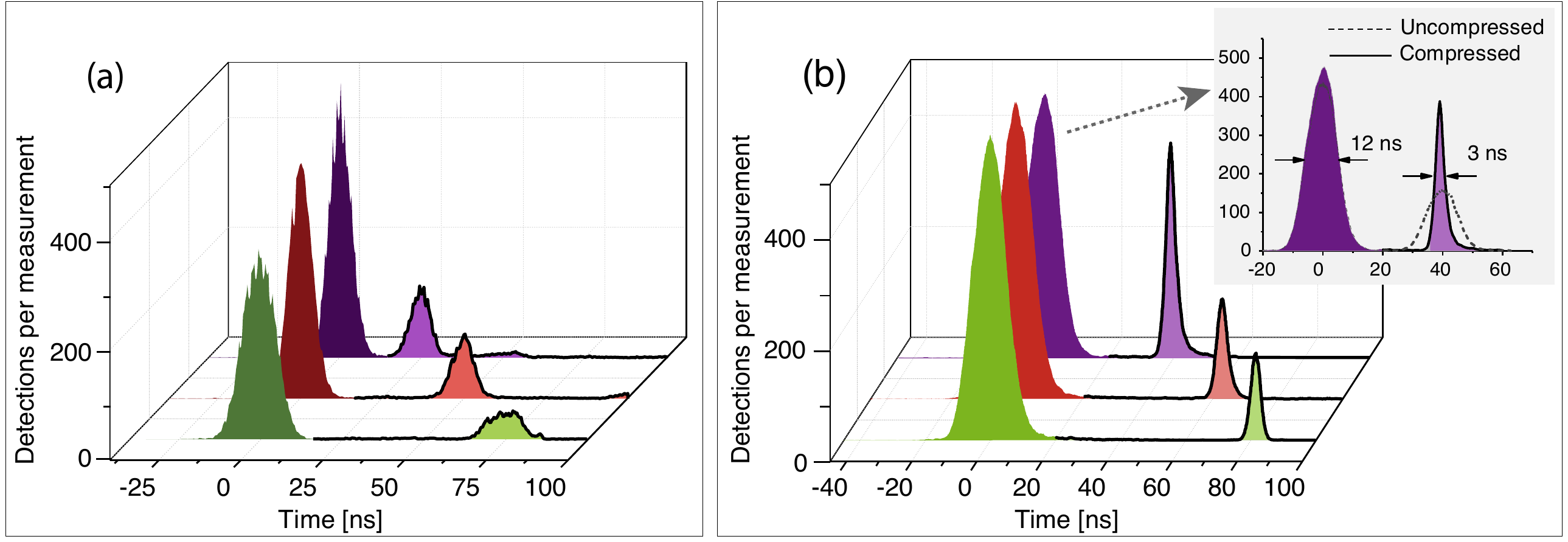} 
\end{center}
\caption[selective delaying]{\textbf{Selective delaying:} \textbf{(a)}~The input phase modulator maps an incoming pulse to a particular segment of the programmed AFC shown in Fig.~\ref{processor} to obtain a desired delay. To achieve delay times of 25 ns, 50 ns and 75 ns, the pulses are mapped onto the segments at -400 MHz, -200 MHz and +200 MHz detuning, respectively, and are retrieved after the corresponding storage time. \textbf{(b)}~A 2 GHz-wide AFC with linearly decreasing peak spacing from 30 MHz to 10 MHz is programmed. The frequency of an incoming pulse is linearly chirped from high to low as described in the previous section. This pulse is mapped onto the programmed AFC, resulting in the emission of a compressed pulse with a compression factor of nearly 4. Varying the start frequency of the chirp  maps the incoming pulse to different intervals of the AFC, which allows selecting the storage time.}
\label{ondemanddelay}
\end{figure}
Having AFC segments with different peak spacing also allows one to select the storage time on demand without the (time demanding) need for re-programming an AFC. In our new approach, the input controller maps a pulse to a particular AFC segment shown in Fig.~\ref{processor} to achieve a desired delay, as demonstrated in Fig.~\ref{ondemanddelay}a. In addition, it is possible to compress the pulses during storage (as shown in the previous section) using an AFC with continuously varying peak spacing. In this case, changing the start frequency of the chirp additionally allows one to choose non-discretized storage times using appropriate frequency shifts at the input controller, as shown in Fig.~\ref{ondemanddelay}b and further described in Appendix 6. If necessary, the emitted pulse's carrier frequency can be set back to the original frequency using the output controller. The described feature will be useful for synchronization purposes in quantum communication. 
 
\section{Discussion and Conclusion} \label{discussion}

An integrated optical quantum memory is an important element for future quantum communication networks and linear optics quantum computers, and employing it for the manipulation of optical pulses at the quantum level adds another direction to its potential use in quantum information processing applications. In this study, we have shown that our broadband waveguide quantum memory can be turned into a processing unit for arbitrary temporal and spectral manipulation of nanosecond-long quantum optical light pulses. Although we utilized weak laser pulses at the few photon level, our approach works equally well for laser pulses containing less than one photon on average, true single photons, or members of entangled pairs of photons. In addition, it can be used for the manipulation of strong pulses of light, as used in classical communications. 

The light-matter interface presented in this paper brings several advantages over previous implementations of quantum optical pulse manipulations. First of all, its waveguide nature allows for integration with other waveguide photonic devices, e.g. currently developed lithium niobate based quantum circuits \cite{Obrien_12}, which is important for practical applications. Second, as our processor relies on a large-bandwidth AFC memory and large-bandwidth electro-optic frequency shifters, it allows processing of nanosecond, and soon sub-nanosecond pulses. Third, as opposed to other quantum memory approaches~\cite{Nunn_08}, the number of simultaneously storable modes in the AFC approach does not depend on the optical depth of the used medium \cite{Afzelius_09, Usmani_10, Thierry_11, Sinclair_13}. Hence, our light-matter interface possesses a large multi-mode processing capacity. And finally, it offers a wide range of quantum optical pulse manipulations in a single spatial mode, which makes it attractive for several quantum information processing applications~\cite{Walmsley_13}.

However, for practical use, the total efficiency of our system (including processor and controllers) needs to be increased beyond the current sub-percent level (for some applications it needs to be close to one). The principal limitation in our current implementation stems from imperfect preparation of the AFC, which is mainly due to unfavorable relaxation dynamics in our material at 3.5 K (see Appendix~A3). More specifically, the residual absorption background of the prepared AFCs, which can be seen in Fig.~\ref{processor}, results in most of the incoming photons to be irreversibly absorbed. However, our recent observations, supplemented by those reported in \cite{Thiel_10}, suggest that the relaxation dynamics change considerably when operating at lower temperature ($< 1.5$ K), and that this background can be eliminated almost entirely. Hence a substantial increase of the efficiency is readily achievable. In addition, to increase the system efficiency, the coupling efficiencies in and out of the phase modulators and our waveguide quantum memory need to be maximized, for instance through integration of all components on a single chip. However, to approach unit efficiency, certain not yet fully mastered techniques need to be applied. One of them requires embedding the AFC medium into an impedance matched cavity \cite{ Sabooni_13, Afzelius_10_2, Moiseev_10_2, Afzelius_14} that features a resonance width compatible with broadband AFCs. Another one is to implement the AFC with  additional spatial periodicity \cite{Tian_13}. Finally, a third option is to reversibly (and with appropriate phase-matching) map the optical coherence that was excited by absorbing a quantum state onto spin states \cite{Afzelius_10, Timoney_12, Gundogan_13}~\footnote{This method requires the availability of an additional spin state. While lacking in thulium, such levels exist in other rare-earth ions.}.  

To conclude, let us note that, while the operation wavelength cannot be chosen freely (it needs to coincide with an atomic transition in a material that allows for persistent hole burning), there is nevertheless some flexibility, in particular as no long coherence time is needed. Of particular interest are Erbium-doped materials \cite{Lauritzen_10, Saglamyurek_14}, due to an absorption line at telecommunication (around 1550 nm) wavelength.

\section{Acknowledgments}

The authors thank W. Sohler, M. George and R. Ricken for providing the Ti:Tm:LiNbO$_3$ waveguide, and M. P. Hedges, C. Simon, J. Jin, C. W. Thiel and W. R. Babbitt for useful discussions and help during the preparation of the manuscript. 
This work is supported by Alberta Innovates Technology Futures (AITF), the National Sciences and Engineering Research Council of Canada (NSERC), the Killam Trusts, and the Carlsberg Foundation.

\section{Appendix}

 \subsection*{A1. Implementation} 
We use a cryogenically cooled Ti:Tm:LiNbO$_3$ channel waveguide quantum memory as a processor in conjunction with two commercial LiNbO$_3$ electro-optic modulators placed before and after the memory. Although in our experiments the LiNbO$_3$ phase modulators are outside the cryogenic system and connected to the input and output of the memory by single mode fibers, all components can, in principle, be implemented on a single chip. This can be achieved by fabricating a long LiNbO$_3$ channel waveguide with Thulium atoms doped only into the central part, and electrodes deposited at the beginning and end of the waveguide. Note that phase modulators similar to ours can be operated at cryogenic temperature \cite{LNatcryo1}. 
\renewcommand{\figurename}{A.}
\subsection*{A2. Material} 
The fabrication of the Tm:LiNbO$_3$ waveguide and basic spectroscopic properties of Tm atoms in this material have already been reported \cite{Sinclair_10}. The waveguide used in the experiments is 10.4 mm long and 3.5 $\mu$m in diameter. The light is butt-coupled into and out of the waveguide using single mode fibers. End-to-end transmission is about 20$\%$. The crystal temperature is kept around 3.5 K during the experiments, which results in nearly 1.5 $\mu$s coherence time on the optical $^3$H$_6\leftrightarrow ^3$H$_4$ transition at 795.5 nm. A magnetic field of 50--130 G, depending on the measurement, is applied along the C$_3$ axis of the crystal in order to activate long-lived nuclear Zeeman spin level, as illustrated in Fig.~A.\ref{leveldiagram}. These levels are used as shelving levels to generate persistent AFCs. 
\setcounter{figure}{0}
\begin{figure}[t!]
\begin{center}
\includegraphics[width=0.4\columnwidth,angle=0]{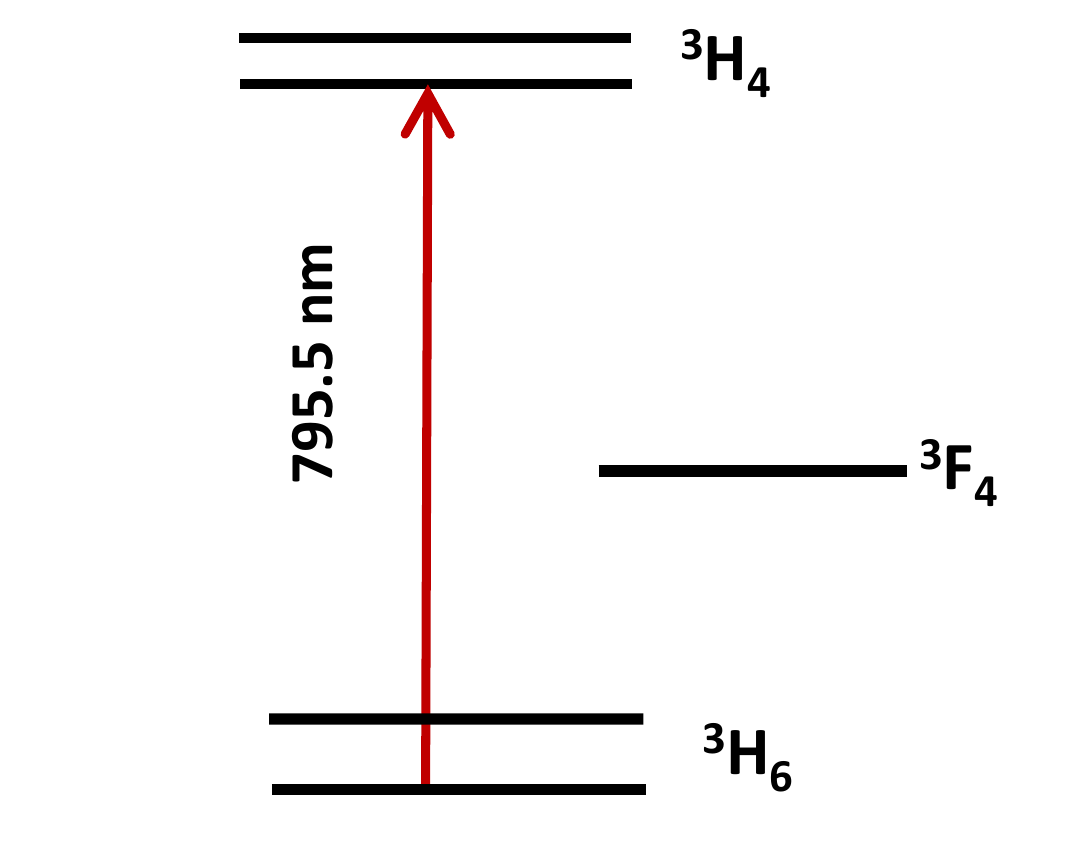} 
\end{center}
\caption{\textbf{Simplified energy level structure of thulium ions:} The optical transition of thulium ions between the ground ($^3$H$_6$) and excited ($^3$H$_6$) electronic level is used for reversible mapping of photonic states through the AFC protocol. The lifetime of the excited level is about $80~\mu s$. Upon application of a magnetic field, the ground and excited levels split into two sub-levels (nuclear Zeeman level) with lifetimes of the two $^3$H$_6$ sub-levels approaching 1~s. These levels are used as shelving states to generate persistent AFC structures through optical pumping described in Appendix~3.}
\label{leveldiagram}
\end{figure}

\subsection*{A3. AFC preparation} Each AFC is prepared by means of frequency selective optical pumping of the inhomogenously broadened transition of Tm at 795.54 nm wavelength. As detailed previously in \cite{Saglamyurek_11}, this is achieved by simultaneously chirping the frequency and modulating the intensity of single frequency laser light. The interaction between the laser light and resonant atoms leads to excitation of the latter. The atoms may subsequently decay to the shelving level. Repeating this processes for a sufficiently long time leads to persistent spectral holes, which form the troughs of the AFC. The atoms that are not excited by the laser are left in the ground level and form the regions of high absorption (the peaks) of the AFC. Adjusting the frequency spacing $\Delta$ between the peaks (or troughs) allows setting the storage time $t_{storage}$ to $t_{storage}=1/\Delta$. 

An important point is that in the optical pumping process, reduced absorption at certain frequencies always comes with increased absorption at other frequencies, as determined by the level structure. In our case, pumping atoms from one magnetic sub-level to the other magnetic sub-level results in decreased and increased absorption at frequencies whose separation is given by the difference between the excited and ground level splittings. Hence we have to adjust the magnetic field in such a way that the increased absorption and reduced absorption regions match the AFC peaks and troughs. However, this approach severely limits experiments that require different peak spacings in the prepared AFCs. Therefore, in some experimental configurations, the magnetic field is optimized so that we obtain nearly equal efficiencies from each AFC segment, at the expense of reducing the overall memory efficiency. Typically, the efficiency was reduced by a factor 2--5 from the optimum efficiency of about 0.5--2~$\%$ for a particular experiment depending on the storage time programmed.

As explained in \cite{Saglamyurek_11} and discussed in Section~\ref{discussion} of the main text, the low efficiencies and relatively short storage time of our memory implementation are in general due to unfavorable relaxation dynamics (arising from low branching ratio and insufficient persistent hole lifetime) and a high decoherence rate in our specific sample kept at 3.5 K. On the other hand, the results of our recent spectroscopic characterizations carried out at temperatures below 1.5 K indicate that substantial improvement in the efficiency and storage time is possible; this is  supported by the observations reported in \cite{Thiel_10}. In addition, improving the fabrication of the Tm doped waveguide will also lead to better spectroscopic properties, as discussed in \cite{Thiel_12}. 

\subsection*{A4. Measurements} In each experimental cycle, optical pumping lasts for 3 ms. After the preparation of an AFC, a 2.2 ms wait time is set to eliminate any fluorescence  from excited atoms, which may swamp the re-emitted pulses. During the next 5 ms, 12 ns long probe pulses are generated at a 2.7 MHz rate. The pulses are attenuated down to a mean photon number of 5-40 and then sent into the AFC memory. As depicted in Fig.~\ref{pulsemanipsetup}, the directly transmitted, and the stored and retrieved pulses are detected by a Si-APD single photon detector with $70\%$ detection efficiency and dark count rate of approximately 100 Hz. Given the small recall efficiency, the estimated mean photon numbers behind the memory are between 0.007 and 0.04 photons per pulse, depending on the experiment. To ensure sufficient detection statistics, each measurement is performed over $0.5-2.5$ minutes. In order to time-resolve and process the detection events, a time-to-digital convertor is used. The optical pumping light with 5 to 15 $\mu$W peak power is sent in backward direction to the memory (i.e. counter-propagating with respect to the probe pulses) to protect the single photon detector. Finally, an optical switch allows toggling the memory output between optical pumping and transmission towards the detector. 

\subsection*{A5. Spectral Manipulation of Pulses Retrieved from the Memory }
 
\begin{figure}[t!]
\begin{center}
\includegraphics[width=1\columnwidth,angle=0]{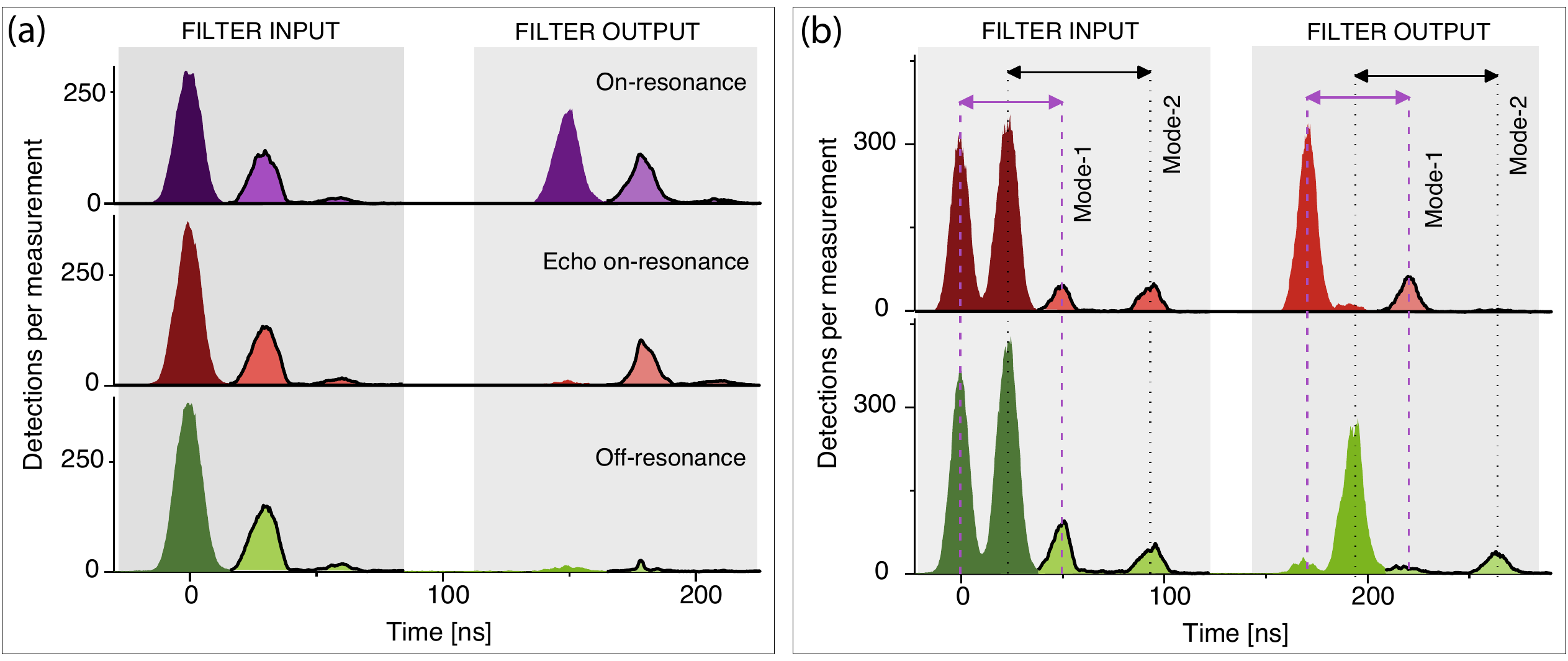} 
\end{center}
\caption[Demonstration of spectral manipulation of  pulses retrieved from the memory unit using the output phase modulator.]{\textbf{Demonstration of spectral manipulation of  pulses retrieved from the memory unit using the output phase modulator:} \textbf{(a)} Pulses with 12 ns duration are stored in an AFC prepared with 600 MHz bandwidth and 33 MHz peak spacing. They are retrieved after 30 ns as shown on the left. The retrieved pulses are directed to the output phase modulator before going to the FP filter. The transmission window of the filter is tuned to the center of the AFC. First, no frequency shift is applied to the transmitted pulse or the echo, and both thus pass the filter, as shown in the top figure. Next, a 200 MHz frequency shift is applied to the transmitted pulse, resulting in only the echo passing the filter, as shown in the middle figure. Finally, a 200 MHz frequency shift applied to both pulses, and thus both are blocked by the filter. \textbf{(b)} Two 25 ns-separated input pulses, occupying different frequency modes, are simultaneously stored in a two-segment AFC with 50 ns and 70 ns storage times, respectively, as shown in the left part of the figure. After recall from the memory, by applying an appropriate frequency shift, we selectively retrieve one of these modes through the filter as shown in the right part.}
\label{FPcavity}
\end{figure}
In most of the demonstrations reported in the main text, the center frequency or spectral distribution of the pulses are changed by the input phase modulator. The main role of the phase modulator that follows the memory unit is to undo these changes, if necessary. For example, when pulses with the same carrier frequency are re-ordered in time, their center frequencies are automatically shifted by certain amounts as explained in Section \ref {pulsereorder} of the main text. In this case, using the output phase modulator, the frequency of the re-ordered pulses can be shifted back to their original values. 

To demonstrate this capability, we add a FP filter with 80 MHz linewidth and 23 GHz free-spectral range behind the output phase modulator (see \cite{Lvovsky_12} for more details about the FP filter used). We generate an AFC with 600 MHz bandwidth and 33 MHz peak spacing, leading to 30 ns storage time, and tune the transmission window of the filter to the center of the AFC. First we store pulses with 12 ns duration in the prepared AFC, and detect the recalled photons after having set the frequency shift introduced by the output phase modulator to zero, as shown at the top of Fig.~A.\ref{FPcavity}a. Next we shift the frequency of the transmitted pulse by -200 MHz but leave the recalled pulse's frequency unchanged. In this case, while the transmitted pulse is blocked by the filter, the recalled pulse passes, as seen in the middle of Fig.~A.\ref{FPcavity}a. Finally we shift the frequency for transmitted and recalled pulses by 200 MHz, causing both to be rejected by the filter as shown at the bottom of Fig.~A.\ref{FPcavity}a.

In another demonstration, we prepare an AFC with two frequency segments, each having 200 MHz bandwidth centered at -100 MHz and +100 MHz detuning. We set the storage times to 50 ns and 70 ns, respectively. Next, we generate two 25 ns separated pulses with -100 MHz and +100 MHz detuning and store them simultaneously in the prepared AFC. They are retrieved after their respective storage times, as shown on the left of Fig.~A.\ref{FPcavity}b. We direct them to the phase modulator and the filter tuned to -100 MHz. Shifting the frequency by 0 MHz and +200 MHz, we can limit the output from the cavity to a single frequency mode, as illustrated on the right hand side of Fig.~A.\ref{FPcavity}b. 

These experiments demonstrate that using the output phase modulator, we have full control over the processed pulses in the frequency domain. In particular, any temporal manipulation that requires the spectra of the pulses to remain unchanged, or any spectral re-distribution that protects the encoded coherent information, can be achieved with the use of this component. Moreover, as shown in \cite{Sinclair_13}, integrating a FP cavity into our system allows for on-demand recall with respect to the frequency domain.

\subsection*{A6. Theoretical Analysis of Pulse Compressing and Stretching} \label{theory} In this section we present a theoretical analysis for compressing and stretching optical pulses using an AFC processor. As described in Section \ref{compress} in the main text, the frequency of an input pulse is linearly chirped and then the resulting pulse is mapped onto an AFC with a linearly varying peak spacing (Fig. \ref{compression1}a), giving rise to a storage time gradient $\mu$. After a predetermined storage time, either a compressed or a stretched pulse emerges, depending on the applied chirp.

We describe the frequency-chirped pulse, propagating along the $z$-direction, before mapping onto the AFC as 
\setcounter{equation}{0}
\renewcommand{\theequation}{A\arabic{equation}}
\begin{equation}\label{Echirp}
E(z=0,t)=e(t)C(t)
\end{equation}
\noindent Here, $e(t)$ is the waveform of the input pulse before chirping and $C(t)=e^{i\omega_1t-irt^2/2}$ is the applied chirp function, where $r$ is the chirp rate and $\omega_1$ is the start (offset) frequency of the chirp \cite{Linget_13}. 

Following the treatment carried out in \cite{Linget_13}, the first echo that is emitted from the medium of length $L$ is expressed by 
\begin{equation}\label{Eout_freq} 
{\tilde E}(z=L,\omega)=C_1e^{-i\mu(\omega-\omega_0)^2/2}e^{-ikL} {\tilde E}(0,\omega), 
\end{equation} 
\noindent where ${\tilde E}(0,\omega)=\int{dt E(z=0,t)e^{-i\omega t}}$ is the Fourier transform of the chirped input pulse. Furthermore, $\omega_0$  is the start frequency of the programmed AFC, and $|C_1|^2$ describes the efficiency of the retrieval. $|C_1|^2$ is determined by the characteristic parameters of the programmed AFC such as optical depth, the finesse and the shape of the AFC peaks but independent of the compression factor. Optimizing these parameters allows in principle approaching unit efficiency in the compression (stretching) process \cite{Linget_13}. As seen in Eq.~\ref{Eout_freq}, the AFC introduces a quadratic phase modulation at rate  $\mu$, i.e. the storage time gradient of the programmed AFC: 
\begin{equation}\label{mu} \mu=\frac{\delta T}{B}. 
\end{equation}
\noindent Here $\delta T=t_{storage}^{max}-t_{storage}^{min}$ describes the difference between the maximum and minimum storage time (recall that the storage time in an AFC memory is given by the inverse tooth spacing) over the bandwidth of the chirped pulse, $B$. 

The output waveform in the time domain can be found using the Fourier transform, $E(L,t)=\frac{1}{2\pi}\int{d\omega e^{i\omega t}{\tilde E}(L,\omega)}$. We assume that the input waveform (before chirping) is described by a Gaussian function of the form $e(t)=e^{-\frac{(t-t_0)^2}{2\tau_{in}^2}}$, where the Full Width at Half Maximum (FWHM) of the intensity is $2\tau_{in}\sqrt{ln(2)}$. Setting $t_0=0$, a lengthy but straightforward calculation yields 
\begin{equation}
|E(L,t)|^2\propto e^{\frac{-(t-t')^2}{\tau_{in}^2\Big 
(\frac{\mu^2}{\tau_{in}^4}+(\mu r-1)^2\Big )}},
\label{output}
\end{equation}

\noindent where $t'=2\mu\omega_1-\mu\omega_0$ is the average storage time. Note that we dropped the factor that describes the efficiency as we are only interested in the temporal shape of the recalled pulses.  Eq.~\ref{output} allows simulating the experimental results as shown in Figs~\ref{compression1} and \ref{highcompression} in the main article. Hence we can calculate the expected duration of the recalled pulse $\tau_{out}=\tau_{in}\sqrt{\frac{\mu^2}{\tau_{in}^4}+(\mu r-1)^2}$, and, after comparison with the duration of the input pulse $\tau_{in}$, we can predict the compression factor $\kappa= \tau_{in}/ \tau_{out}$. Note that adjusting the start frequency of the chirp ($\omega_1$) with respect to the start frequency of the AFC ($\omega_0$) allows setting the storage time, as demonstrated in Section~\ref{ondemand}. 

As discussed in the main text, the optimal compression factor is obtained for $\mu r=1$ (i.e. $\tau_{in}=\delta T$ ). Under this condition, the input pulse $|e(t)|^2=e^{-\frac{t^2}{\tau_{in}^2}}$ is converted into the output pulse $|E(L,t)|^2=e^{\frac{-(t-t')^2}{\tau_{out}^2}}$, where $\tau_{out}=\frac{\mu}{\tau_{in}}$ (from Eq. \ref{output}), allowing us to calculate the compression factor $\kappa=\frac{\tau_{in}^2}{\mu}$. This expression shows that higher compression factors can be achieved by extending the duration of the input pulse, as demonstrated in Fig.~\ref{highcompression}. Note that it is equivalent to $\kappa=\frac{\delta T^2}{\mu}=\mu B^2$, as described in Section~\ref{compress}.
  
Let us finally point out that the possibility for active control of the parameter $\mu r$ by adjusting the chirp rate allows one to implement compressing and stretching in a very flexible fashion. As detailed in \cite{Linget_13}, having a value of $\mu r$ between 0 and 1 leads to compression (with variable compression factor) of the input pulse without changing the order of the raising and falling edge, while setting $\mu r$ between 1 and 2 causes a compressed, but time-reversed output. If $\mu r$ is equal to 2,  time reversal of the input pulse without compression is obtained. On the other hand, setting $\mu r$ larger than 2 leads to a time-reversed stretched pulse. In addition, flipping the sign of the chirp, as described in the main text, allows one to implement stretching without time-reversal. In this case a larger $\mu r$ results in increased  stretching. Combining these features with the large time-bandwidth product of the AFC and the large bandwidth of existing phase modulators makes it possible to change the duration of pulses in a fully controlled manner by several orders of magnitude. 

 \subsection*{A7. Serrodyne Frequency Shifting} 
The serrodyne frequency shift technique \cite{Serrodyne_1} is an essential tool used in our experiments;  it was employed in most of the described pulse manipulations and AFC preparations. The principle of this technique is to introduce a linearly varying time-dependent phase change to the light by applying a sawtooth shaped voltage to an electro-optic phase modulator. As a result of the linear phase change, the frequency of the light is shifted by the amount of the modulation frequency. The great advantage of this technique is that, in principle, all energy can be transferred to the desired frequency mode without producing side-bands. This feature allows high-bandwidth spectral manipulations of short optical pulses, which we exploited for our experiments. To generate a sawtooth modulated voltage, we use an arbitrary waveform generator (AWG) with 20 GS/s sampling rate. The output voltage of the AWG is amplified using a high-speed amplifier to achieve the $2\pi$-voltage value of the electro-optic modulators which is necessary for providing maximum energy transfer to the desired frequency mode. Adjusting the modulation voltage between zero and the $2\pi$-voltage allows us to vary the amount of energy transferred from the carrier frequency to the desired side-band modulation. Our commercial electro-optic phase modulators have 20 GHz bandwidth and feature approximately 2 dB loss. We are able to shift the light frequency by up to $\pm 5$ GHz. Having imperfections in the applied sawtooth waveform causes imperfect energy transfer efficiency to the desired frequency mode. For instance, for a frequency shift of about 1 GHz, we maximally transfer 80$\%$ of the unmodulated light's power.

\end{document}